\documentclass[prd,preprintnumbers,nofootinbib,aps,9pt]{revtex4}
\usepackage{subeqnarray}
\usepackage{epsfig,amsmath,amssymb}
\usepackage{mathrsfs}
\usepackage[usenames,dvipsnames]{color}
\usepackage[pagebackref=true, colorlinks=true]{hyperref}
\definecolor{redish}{rgb}{0.7,0.2,0.0}  
\definecolor{bluish}{rgb}{0.2,0.5,0.8}
\hypersetup{linkcolor=redish,          
                  citecolor=blue,        
                  filecolor=magenta,      
                  urlcolor=bluish}          

\DeclareFontFamily{U}{rsfs}{}         
\DeclareFontShape{U}{rsfs}{m}{n}{<5> rsfs5 <6><7> rsfs7          %
  <8><9><10><10.95><12><14.4><17.28><20.74><24.88> rsfs10}{}     %
\DeclareMathAlphabet{\mathfs}{U}{rsfs}{m}{n}                     %
\newcommand{\mfs}[1]{\mathfs {#1}}                              %

\newcommand{\ba}{\nopagebreak[3]\begin{eqnarray}}
\newcommand{\ea}{\end{eqnarray}}
\newcommand{\bii}{\begin{itemize}}
\newcommand{\eii}{\end{itemize}}
\newcommand{\nn}{\nonumber}

\newcommand{\sO}{{\mfs O}}

\newcommand{\f}{\frac}

\def \l{\ell}

\def \lp{\l_p}

\def \th{\theta}

\def \O{\Omega}

\def \({\left(}
\def \){\right)}
\def \[{\left[}
\def \]{\right]}

\def\pb#1{\rlap{\lower1.5ex\hbox{$\longleftarrow$}}{#1}}
\def\dpb#1{\rlap{\lower1.5ex\hbox{$\Longleftarrow$}}{#1}}
\def\spb#1{\rlap{\lower1.5ex\hbox{$\leftarrow$}}{#1}}
\def\sdpb#1{\rlap{\lower1.5ex\hbox{$\Leftarrow$}}{#1}}
\begin{document}
\title{Note on $SU(2)$ isolated horizon}
\author{Abhishek Majhi}%
\email{abhishek.majhi@gmail.com}
\affiliation{Indian Statistical Institute,\\Plot No. 203, Barrackpore,  Trunk Road,\\ Baranagar, Kolkata 700108, West Bengal, India\\}
\begin{abstract}
  
  We point out that the symplectic structure, written in terms of the Sen-Ashtekar-Immirzi-Barbero variables, of a spacetime admitting an isolated horizon  as the inner boundary, involves a positive  constant parameter, say $\sigma$, if $\gamma\neq\pm i$, where $\gamma$ is the Barbero-Immirzi parameter. The parameter $\sigma$ represents the rescaling freedom that characterizes the equivalence class of null generators of the isolated horizon. {We reiterate the fact that} the  laws of mechanics  associated with the isolated horizon does not depend on the choice of $\sigma$ { and, in particular}, while one uses the value of standard surface gravity as input, that does not fix $\sigma$ to a particular value.  { This fact contradicts the claims made  in certain parts of the concerned literature that we duly refer to. We do  the calculations by taking Schwarzschild metric as an example so that the contradiction with the referred literature, where similar approaches were adopted, becomes apparent.} { The contribution to the symplectic structure that comes from the isolated horizon}, diverges for $\sigma^2=(1+\gamma^2)^{-1}$, implying that the rescaling symmetry of the isolated horizon is violated for any real $\gamma$. Since the quantum theory of $SU(2)$ isolated horizon in the LQG framework exists only for real values of $\gamma$, it is founded on this  flawed classical setup. { Nevertheless, if the flaw is ignored, then two different viewpoints persist in the literature for entropy calculation. We highlight the main features of those approaches and point out why one is logically viable and the other is not.}

\end{abstract}
\maketitle
\section{Introduction}
The canonical quantization program of gravity, known as loop quantum gravity (LQG), is based on real $SU(2)$ gauge fields on a spatial slice, called Sen-Ashtekar-Immirzi-Barbero (SAIB) variable \footnote{See \cite{thiemann} for the historical reasons behind this nomenclature.} and its conjugate momentum \cite{thiemann,ashlew}. The SAIB connection involves a free parameter called Barbero-Immirzi parameter 
$(\gamma)$\cite{imm1,imm2,bar}. Although the introduction of such variables has provided us with a viable quantum theory for real and positive values of $\gamma$, nevertheless, it is a very well known fact that the SAIB gauge field acts as a connection only on a spatial slice; it can not be interpreted as the pullback of a spacetime connection on a slice unless $\gamma=\pm i$ \cite{samuel}. In spite of this feature of the SAIB connection, the available theory of LQG has been founded on the SAIB connection only for real and positive values of $\gamma$ \cite{ashlew,thiemann}, which has set the stage for black hole entropy calculation \cite{qg1,qg2,km98,km00,sigma,per1,per2,per3,per4, bhentropy}.
 The novelty of the framework lies in the fact that it gives a clear path from the classical to the quantum theory, leading to the Hilbert space structure of the black hole horizon and hence, counting of states. This enables one to calculate black hole entropy from the first principles using statistical mechanics.

Some efforts have been made to address the issue of black hole entropy calculation for $\gamma=\pm i$ \cite{perezcomplex, karimcomplex}. However, those calculations are mainly focussed on obtaining the Bekenstein-Hawking area law \cite{bek,haw} from the already available  results for real and positive values of $\gamma$, by using mathematical techniques such as analytic continuation \cite{perezcomplex,karimcomplex}. Unlike the real-$\gamma$ scenario, there is no  derivation of black hole entropy from the first principles, whose beginning is rooted to the classical theory. In other words, for $\gamma=\pm i$ there is not yet satisfactory answers to the more fundamental questions that arise before one talks about entropy calculation, like what are the quantum states on the black hole horizon, what is the associated Hilbert space, how do we count the states, etc.

Now, a black hole horizon in equilibrium, is modeled as an isolated horizon (IH), which is taken to be a null inner boundary of the spacetime satisfying  certain boundary conditions \cite{ih0,ih1,ih2,ih3}. The theory of IH is based on either real $SO(1,3)$ or complex $SL(2,C)$ {\it spacetime} connections\cite{ih0,ih1,ih2,ih3}. { Importantly, the laws of mechanics follow from the boundary conditions that define the IH and, as well, without violating any of its intrinsic symmetries. This has been elaborately explained with full generality, especially, in \cite{ih3}.} In \cite{ih1}, the action of a spacetime with IH as an inner boundary was written, from which the symplectic structure (on a slice) was deduced. {\it It may be noted that until this point it was a calculation with spacetime connection.}

Right at this stage, {\it real} $SU(2)$ SAIB variable, which behave as a connection only on a spatial slice, was introduced. { Digressing a bit we may note that, as far as our knowledge is concerned, no proof of the laws of mechanics for the IH with such a variable, especially along the lines of investigation explored in \cite{ih3}, exists in the literature till date. To be even more critical, we may further note that there is no construction of the whole IH framework available, in the literature till date, that is founded on such a variable. The obstacle for performing such analyses is the  basic fact that such a variable can not be interpreted as the pullback of a spacetime connection and therefore, right at the outset it is just a mere {\it assumption} that we are still doing a theory of IH with such a variable. After all, an IH is just a local construction that generalizes the global notion of event horizon of a black hole {\it spacetime.} Nonetheless, such obstacles have remained ignored.}

 Now, the passage from $SL(2,C)$ connection to the spatial real SAIB variable involve the following steps: i) pullback of the spacetime connection to a spatial slice is considered ii) an internal vector is kept fixed and only the rotations, which keep this internal vector invariant, are allowed, therefore, reducing the internal gauge group to a complex $SU(2)$ iii) the complex  variables are made real by replacing the $i$ by a real parameter $\gamma$ according to the prescription suggested in \cite{bar}. Finally, one has a real $SU(2)$ connection on a slice i.e. the SAIB  connection. 

Importantly, the full $SU(2)$ gauge group was further reduced to $U(1)$ on the IH, although the authors {\it admitted} that {the gauge fixing was unnecessary} (as it should be if the theory has to have a physical interpretation at all)\cite{ih1}. Therefore, the symplectic structure of a spacetime admitting IH, with the full $SU(2)$ gauge group was not manifest. In the mean time, based on the available scenario, it was proposed in \cite{km98}, that the full gauge group on the IH should be $SU(2)$ in the context of black hole entropy calculation from the quantum theory. However, the derivation of the symplectic structure at the classical level with the full $SU(2)$ group on the IH remained pending. This was accomplished, about a decade after those earlier works, in \cite{per2}. Notably, it came with an interesting twist. In the symplectic structure for the $SU(2)$ case \cite{per2}, the prefactor in front of the contribution from the IH, differed from that of the $U(1)$ case \cite{ih1}. While in the $U(1)$ scenario it was only the area of the IH (say, $A_{IH}$) \cite{ih1}, in the $SU(2)$ case it was  $A_{IH}/(1-\gamma^2)$\cite{per2}.

{ Now, a careful reading of ref.\cite{per2}  reveals that the value of a certain  parameter was fixed in order to obtain the factor $A_{IH}/(1-\gamma^2)$, by arguing that the fixation is the only choice if one wants to have the correct value of the surface gravity. Such an argument is, at the least, ill-founded because, as we have pointed out earlier, no construction of the IH horizon framework and proof of the laws of IH mechanics exist in the literature starting from the real $SU(2)$ SAIB variables and therefore, it is not even known how to make sense of the word `surface gravity'. Our motive is to discuss the elementary details of the sort of calculations that led to the conclusions drawn in ref.\cite{per2} and pin point how the `surface gravity' argument does not hold.
	
	In view of this}, we investigate a specific example of a spacetime admitting an $SU(2)$ IH as an inner boundary and show that the prefactor in front of the contribution from the IH actually contains a parameter ambiguity, say $\sigma$. This ambiguity is exactly the one which is present in the choice of null generators of the IH; the null-generator of an IH is unique up to a positive constant \cite{ih0}. We check explicitly that the zeroth law, the first law and the value of surface gravity associated with the IH does not depend on the choice of the parameter\footnote{This is just a verification, with a particular simple and concrete example, of what is already evident in the general framework investigated in \cite{ih3}.}. This parametric ambiguity is absent for $\gamma=\pm i$ when the SAIB connection has a spacetime interpretation. Also, { the contribution to the symplectic structure that comes from the isolated horizon}, diverges for $\sigma^2=(1+\gamma^2)^{-1}$ implying that the full symmetry of an $SU(2)$ IH is not retained. Above that, any other choice of a particular value of the parameter, like the one made in ref. \cite{per2},  provides a fixed relationship between the local Lorentz boosted frame and the foliation of the IH which violates the symmetries of an IH even if we restrict to the condition  $\sigma^2\neq(1+\gamma^2)^{-1}$. Taking all these facts into considerations, we conclude that the quantum theory of $SU(2)$ isolated horizon in LQG  framework is founded on a flawed classical setup { that is devoid of a rigorous derivation of the laws of IH mechanics and consequently	fails to motivate the necessity of an understanding of quantum states and entropy calculation in the first place.}

{ Nonetheless, if the flaw is ignored, then two different viewpoints persist in the literature concerned with the entropy calculation in the $SU(2)$ framework. One can be found in refs.\cite{sigma,km11} and other can be found in ref.\cite{per4}. We highlight the features of the two viewpoints in the context of the present analysis and by taking into consideration the consequences of an experimental determination of $\gamma$. We point out that, in such a case, while the approach taken in refs.\cite{sigma,km11} can only be falsified like any other theory having experimentally fitted parameters, the approach taken in ref.\cite{per4} leads to a relation between $A_{IH}$ and $\sigma$ that is suggestive of the fact that for a given area there is a preferred Lorentz boosted frame on the horizon! Hence, we conclude that the approach of ref.\cite{km11,sigma} is logically viable and that of ref.\cite{per4} is not.}

The structure of the paper can be debriefed as follows. In section \ref{section1} we consider a specific example of a spherically symmetric IH and explicitly point out the fact that the zeroth and the first law of mechanics and the value of surface gravity associated with the IH are independent of the parameter, $\sigma$, that represents the rescaling freedom of the null generators of the IH. In section \ref{section2} we investigate the necessary equations to write down the symplectic structure and point out the relevant issues concerning the appearance of the parameter $\sigma$ in the symplectic structure of the IH written in terms of SAIB variables. { We conclude that the quantization program of $SU(2)$ IH is based on a flawed classical setup.} { In section \ref{remarks} we highlight the features of the entropy calculation performed in refs.\cite{per2,per4,sigma,km11} to point out that ref.\cite{per2,per4} showcase self-contradictory research and we draw further appropriate conclusions regarding the approaches taken in refs.\cite{km11,sigma} and  ref.\cite{per4}.} { Finally, in section \ref{section3} we end with some concluding remarks after providing a summary of this work.} 



\section{Physical laws governing the black hole horizon}\label{section1}
In this section, we shall consider a portion of the Schwarzschild spacetime as a simple and concrete example of the more general quasi-local framework of non-expanding horizon and isolated horizon \cite{ih0}. We investigate the implications of the laws of mechanics on the horizon.
\subsection{Section of a Schwarzschild spacetime}
Let us consider a section $\cal M$ of a spacetime 
described by the Schwarzschild metric (see e.g. \cite{inverno}):
\begin{eqnarray}
ds^2=g_{\mu\nu}dx^{\mu}dx^{\nu}=\O(t,x)(-dt^2+dx^2)+r^2(t,x)~d\th^2+r^2(t,x)\sin^2\th~d\phi^2\label{metric}
\end{eqnarray}
with the bound $r(t,x)\geq 2M$, where 
\begin{eqnarray}
\O(t,x)=\frac{16 M^2}{r(t,x)}\exp[-r(t,x)/2M]
\end{eqnarray}
subject to
\begin{eqnarray}
&&t^2-x^2=-(r-2M)e^{r/2M},
\end{eqnarray}
which implies
\begin{eqnarray}
dr=\f{4M}{r}~e^{-r/2M}(x~dx-t~dt).
\end{eqnarray}
$\cal M$ has an inner null boundary $\Delta$ at $r=2M$, which is a spherically symmetric, uncharged non-expanding horizon (NEH) with constant area $A_{\Delta}=16\pi M^2$ of cross-section \cite{ih0,ih1,ih2,ih3}. $\cal M$ admits a time-like Killing field $\chi$ (e.g. see \cite{wald}). Let us choose a time-function $\Phi(t,x,\th,\phi)$, such that the time evolution vector field $T:=d/d\Phi$ satisfies the condition $\spb{T}=\spb{~\chi}$; we denote any quantity $Q$ pulled back to $\Delta$ is denoted by $\spb{~Q}$. 
$\cal M$ is bounded `initially' and `finally' by the spatial hypersurfaces given by $\Phi=\Phi_i$ and $\Phi=\Phi_f$ respectively.

The non-zero components of the metric are as follows:
\begin{eqnarray}
g_{tt}=-\Omega ,~ g_{xx}= \Omega ,~g_{\th\th}= r^2,~  g_{\phi\phi}=r^2 \sin ^2\theta. \label{mc}
\end{eqnarray}
The non-zero components of the inverse of the metric are as follows:
\begin{eqnarray}
g^{tt}=-\Omega^{-1},~g^{xx}={\Omega^{-1}},~g^{\th\th}= r^{-2},~g^{\phi\phi}=r^{-2} \csc ^2\theta. \label{imc}
\end{eqnarray}
The non-zero components of the Christoffel symbol 
$\Gamma^\mu_{\alpha\beta}:=\frac{1}{2}g^{\mu\sigma}\left(\partial_{\alpha}g_{\sigma\beta}+\partial_{\beta}g_{\sigma\alpha}-\partial_{\sigma}g_{\alpha\beta}\right),$
 are as follows:
\begin{eqnarray}
&& \text{$\Gamma^t_{~tt} $}=  \frac{\dot\Omega }{2 \Omega},~~~~~~~
 \text{$\Gamma^t_{~xt} $}=  \frac{\Omega^{\prime}}{2 \Omega},~~~~~~~
 \text{$\Gamma^t_{~xx} $}=  \frac{\dot\Omega}{2 \Omega},~~~~~~~ \text{$\Gamma^t_{~\th\th} $}=  \frac{r \dot r}{\Omega},~~~~~~~
 \text{$\Gamma^t_{~\phi\phi} $} = \frac{r \dot r \sin ^2\theta  }{\Omega
   },~\nn\\
   &&
 \text{$\Gamma^x_{~tt} $}=  \frac{\Omega^{\prime}}{2 \Omega},~~~~~~~
 \text{$\Gamma^x_{~xt} $}=  \frac{\dot\Omega}{2 \Omega},~~~~~~~
  \text{$\Gamma^x_{~xx} $}=  \frac{\Omega^{\prime}}{2 \Omega},~~~~~~~
 \text{$\Gamma^x_{~\th\th} $}=  -\frac{r r^{\prime}}{\Omega},~~~~~~~
 \text{$\Gamma^x_{~\phi\phi} $}=  -\frac{r r^{\prime} \sin ^2\theta  }{\Omega
   },\nn\\
   &&
 \text{$\Gamma^{\th}_{~\th t} $}=  \frac{\dot r}{r},~~~~~~~
  \text{$\Gamma^{\th}_{~\th x} $}=  \frac{r^{\prime}}{r},~~~~~
 \text{$\Gamma^{\th}_{~\phi\phi} $}=  -\cos \theta \sin \theta ,~~~~~
 \text{$\Gamma^{\phi}_{~\phi t} $}=  \frac{\dot r}{r} ,~~~~
 \text{$\Gamma^{\phi}_{~\phi x} $}=  \frac{r^{\prime}}{r},~~~~
 \text{$\Gamma^{\phi}_{~\phi\th} $}=  \cot \theta,  ~~~~~~~~~~~~~~\label{chris}
\end{eqnarray}
where 
\begin{eqnarray}
&&\dot{r}:={\partial_t r}=-\frac{4Mt}{r}\exp[-r/2M],~~~~
{r^\prime}:={\partial_x r}=\frac{4Mx}{r}\exp[-r/2M],\nn\\
&&
\dot{\Omega}:=\frac{\partial\Omega}{\partial t}=-\frac{16M^2}{r}\(\frac{1}{r}+\frac{1}{2M}\)\exp[-r/2M]\dot r,~~~~
\Omega^\prime :=\frac{\partial\Omega}{\partial x}=-\frac{16M^2}{r}\(\frac{1}{r}+\frac{1}{2M}\)\exp[-r/2M] r^\prime.~~~\label{someresults1}
\end{eqnarray}

\subsection{Standard definition of surface gravity}
The Schwarzschild metric admits a time-like Killing vector field and in the chosen coordinates the contravariant and covariant components of the time translation Killing vector field are given by
\begin{eqnarray}
\chi^\mu=\f{1}{4M}(x,t,0,0)~~~\text{and}~~~\chi_\mu=\f{1}{4M\Omega}(-x,t,0,0)\label{chi}
\end{eqnarray}
respectively. $\chi$ satisfies the conditions~ $\lim_{r\to\infty}g_{\mu\nu}\chi^{\mu}\chi^{\nu}=-1~\text{and}~\lim_{r\to 2M}g_{\mu\nu}\chi^{\mu}\chi^{\nu}=~0.$ 
Due to the presence of the Killing field, using the standard definition \cite{wald}, the surface gravity associated with $\Delta$ can be calculated to be   
\begin{eqnarray}
\kappa=\lim_{r\to 2M}\frac{[g_{\mu\nu}(\nabla_{\chi}\chi^\mu)(\nabla_{\chi}\chi^\nu)]^{1/2}}{(-\chi_\mu \chi^\mu)^{1/2}}=\frac{1}{4M}.\label{keh}
\end{eqnarray}


\subsubsection{Pullback of the Killing vector field on $\Delta$}
Now, we shall find the pullback of the Killing vector field on $\Delta$. On $\Delta$, i.e. for $r=2M$, we have $t^2=x^2$ which gives us two cases $x=t$ and $x=-t$. We shall work with $x=t$ which is the future directed NEH. Let the intrinsic coordinates of $\Delta$ be $(u,\tilde\theta, \tilde\phi)$. Then we have the following relations between the $\Delta$ coordinates and the spacetime coordinates:
\begin{eqnarray}
t=x=f(u),~~\tilde\theta=\theta,~~\tilde\phi=\phi.
\end{eqnarray}
where $f$ is a positive definite invertible function of $u$ such that $df/du$ is non-vanishing. For simplicity we shall use $\theta,\phi$ instead of $\tilde\theta,\tilde\phi$ as the coordinates $\Delta$. Therefore, the pullback of the Killing vector field on $\Delta$ is
\begin{eqnarray}
\spb{~\chi}=\pb{\chi^\mu\partial_\mu}=\frac{1}{4M}\(\pb{x\partial_t}+\pb{t\partial_x}\)
=\frac{1}{2M}\frac{f(u)}{f'(u)}\partial_u\label{chipullback}
\end{eqnarray}
where $f'(u)=df/du$.

\subsection{Surface gravity of $\Delta$ from quasi-local definition}
To use the quasi-local definition of surface gravity associated with $\Delta$ one can take the following steps \cite{per2}. We construct the tetrad:
\begin{eqnarray}
e^{\bf 0}=\O^{1/2}(\cosh {\bf \alpha} ~dt+\sinh {\bf \alpha}~dx),~e^{\bf 1}=\O^{1/2}(\sinh {\bf \alpha} ~dt+\cosh{\bf \alpha}~dx),~e^{\bf 2}=r~d\th,~e^{\bf 3}=r\sin\th~d\phi\label{ot}
\end{eqnarray}
which are related to the metric via the relation $\eta_{IJ}e^I_\mu e^J_\nu=g_{\mu\nu}$, $\eta_{IJ}$ being the internal flat metric  $diag~(-1,1,1,1)$.
 ${\alpha}$ is an arbitrary function of spacetime coordinates, that characterizes the arbitrariness of the local Lorentz boost. Further, one can construct the following set of null tetrad
\begin{eqnarray}
\l=\frac{1}{\sqrt{2}}(e^{\bf 1}-e^{\bf 0}),~k=-\frac{1}{\sqrt{2}}(e^{\bf 0}+e^{\bf 1}),
~m=\f{1}{\sqrt 2}(e^{\bf 2}+ie^{\bf 3}),~\bar m=\f{1}{\sqrt 2}(e^{\bf 2}-ie^{\bf 3})\label{nt}
\end{eqnarray}
which satisfy  $\l.k=-1=-m.\bar m$ and other contractions vanish. Using eqs.(\ref{ot}) and (\ref{nt}) one can obtain the covariant components of $\l$ and $k$ respectively:
\begin{eqnarray}
\l_{\mu}=\frac{1}{\sqrt{2}}\O^{1/2}(\exp-{\bf \alpha})~(-1,1,0,0),~~~~k_{\mu}=-\frac{1}{\sqrt{2}}\O^{1/2}(\exp{\bf \alpha})~(1,1,0,0).
\end{eqnarray}
The contravariant components of $\l$ and $k$ are as follows:
\begin{eqnarray}
\l^{\mu}=\frac{1}{\sqrt{2}}\O^{-1/2}(\exp-{\bf \alpha})~(1,1,0,0),~~~~k^{\mu}=-\frac{1}{\sqrt{2}}\O^{-1/2}(\exp{\bf \alpha})~(-1,1,0,0).
\end{eqnarray}
One can show that
\begin{eqnarray}
\l^\mu\nabla_\mu \l^\nu=\underbrace{\frac{1}{\sqrt{2}}\Omega^{-1/2}(\exp-{\bf \alpha})\(-{\dot\alpha} -\alpha'+\frac{\Omega^\prime}{2\Omega}+\frac{\dot\Omega}{2\Omega}\)}\l^\nu.
\end{eqnarray}
The under-braced quantity evaluated on (or pulled back to) $\Delta$ is identified as surface gravity of the NEH i.e.
\begin{eqnarray}
\kappa_{\Delta}=\frac{1}{\sqrt{2}}~\spb{\Omega}^{-1/2}[\exp-(\spb{~\alpha})]\[-\spb{\dot\alpha} -\spb{\alpha'}+\frac{1}{2\spb{\Omega}}\(\spb{\Omega^\prime}+\spb{\dot\Omega}\)\].\label{kih}
\end{eqnarray}
It may be noted that ${\spb{~\dot\alpha}}\neq \frac{1}{f'(u)}\partial_u \(\spb{~\alpha}\) \neq\spb{~\alpha'}$ in general. Now, we have the following results: 
\begin{eqnarray}
&&\spb{\Omega}=\frac{8M}{\text e},~~\spb{~\dot{r}}=-\frac{2f(u)}{\text e},~~\spb{~{r^\prime}}=\frac{2f(u)}{\text e},~~\spb{\dot{\Omega}}=\frac{16f(u)}{\text e^2},~~\spb{\Omega^\prime}=-\frac{16f(u)}{\text e^2}.\label{oneh}
\end{eqnarray}
where $\text{e}=\exp[1]$. Using the results of (\ref{oneh}) in eq.({\ref{kih}}) we obtain
\begin{eqnarray}
\kappa_{\Delta}=-\frac{1}{4}\(\frac{\text e}{M}\)^{1/2}\(\exp-\spb{~\alpha}\)\(\spb{\dot\alpha} +\spb{\alpha'}\)\label{surf}.
\end{eqnarray}
The pullback of the null tangent vector field $\l$ can be calculated to be:
\begin{eqnarray}
\spb{~\l}&&=\pb{\l^\mu\partial_\mu}=\pb{\l^t\partial_t}+\pb{\l^x\partial_x}=\spb{~\l^t}\partial_f+\spb{~\l^x}\partial_f=\frac{1}{\sqrt 2}~\spb{\Omega}^{-1/2}\(\exp-\spb{~\alpha}\).~2\partial_f=\frac{1}{2}\(\frac{\text e}{M}\)^{1/2}\frac{\exp-\spb{~\alpha}}{f'(u)}~\partial_u,~~\label{ellpullback}
\end{eqnarray}
where $\spb{~\l^t}$ means the component $\l^t$ evaluated on $\Delta$, etc. and we have to use (\ref{oneh}) to reach the final result.

\subsection{Non-expanding horizon and the zeroth law} 
Zeroth law states that the surface gravity must be a constant on the associated horizon i.e. in the present scenario we need to have $\kappa_{\Delta}$ to remain constant along  $\Delta$. This happens if  the Lie derivative  of $\kappa_{\Delta}$ along $\spb{~\l}$ vanishes \cite{ih3} i.e.  $\pb{\l^\mu\partial_\mu}\kappa_{\Delta}=0$. Considering eq.(\ref{surf}) and eq.(\ref{ellpullback}), this yields
\begin{eqnarray} 
\spb{\dot\alpha}+\spb{\alpha'}=m_0\exp\spb{~\alpha}\label{zero}
\end{eqnarray}
where $m_0$ is some negative definite quantity on $\Delta$ such that $\pb{\ell^\mu\partial_{\mu}m_0}=0$. It should be noted that we have considered functional rescaling of $\spb{~\ell}$, allowing $\spb{~\alpha}$ to be a function on $\Delta$ according to the definition of an NEH \cite{ih3}. This means that the {\it zeroth law is valid for $\Delta$, an NEH}, if the rescaling function obeys eq.(\ref{zero}) .     

\subsection{Isolated horizon and the first law}
 An isolated horizon (IH) \cite{ih0,ih1,ih2,ih3}, by definition, only allows positive constant rescaling of $\spb{~\ell}$. In the present analysis, we can see that this is tantamount to tying the foliation of $\Delta$, characterized by $f(u)$, to the local Lorentz boosts on it, characterized by $\spb{~\alpha}$, in the following particular fashion:
\begin{eqnarray}
&&f(u)=\tilde c\exp-\spb{~\alpha},~~~  \text{with $\tilde c >0$ and $\pb{\ell^\mu\partial_\mu}~\tilde c=0$}.\label{ihcondition}
\end{eqnarray}
Considering eq.(\ref{chipullback}) and eq.(\ref{ellpullback}), the condition (\ref{ihcondition}) implies 
\begin{eqnarray}
&& \spb{~\ell}=c_0\spb{~\chi},~~~ \text{with $c_0= (M\text{e})^{1/2}/\tilde c$.} \label{ihcondition1}
\end{eqnarray}  
It is only an IH, {\it not} an NEH, that is associated with a local first law in order to have a Hamiltonian evolution in the covariant phase space \cite{ih3}. Therefore, the surface gravity associated with the IH is $\kappa_{\Delta}$ subject to the condition (\ref{ihcondition}), which, using eq.(\ref{kih}) and eq.(\ref{zero}), gives 
$\kappa_{IH}=\beta \kappa$ with  $\beta:=-m_0(M\text{e})^{1/2}$. Therefore, there is an ambiguity in the exact form of surface gravity that remains in the local first law associated with the IH \cite{ih3}:
 \begin{equation}
 \delta E_{IH}=\frac{\beta\kappa}{8\pi}~\delta A_{IH}.
 \end{equation} 
where $E_{IH}$ is the local energy associated with the IH. In general one has the radiation energy exterior to the IH given by $E_{radiation}=E_{IH}-E_{ADM}$ \cite{ih3}. In the present scenario we have $E_{radiation}=0$ which provides $E_{IH}=E_{ADM}=M$. This fixes the value $\beta=1$ (as $A_{IH}=16\pi M^2$), which implies $m_0=-(M\text{e})^{-1/2}$ (see \cite{ih3} for a more detailed discussion). {\it Importantly, $c_0$ remains arbitrary.} This physically means that there is an equivalence class of $\spb{~\ell}$ related to $\spb{~\chi}$ up to a positive constant,  that leads to the value of the surface gravity of the IH to be $\kappa_{IH}=\kappa=1/4M$. An alternative physical meaning can be extracted by looking at eq.(\ref{ihcondition}). It implies that the foliation of IH, characterized by $f(u)$, and the local Lorentz boosted frame on the IH, characterized by $\spb{~\alpha}$, have a relationship only up to a positive constant ambiguity.

 \subsubsection{A redundant choice: $c_0=1$} \label{subred}
 { At this point, let us precisely point out a fact regarding ref.\cite{per2} because this work has been a crucial founding stone in further development of the concerned literature \cite{perezcite}.} 
 
 It should be noted that for the zeroth law and the first law to hold on the IH, even to fix the value of $\kappa_{IH}$, we {\it do not}  require to choose a particular value of $c_0$.
{ This is expected because $c_0$ is nothing but a representative of the rescaling freedom of the null generators of the IH and the local Lorentz boost and the foliation of the IH are fixed only up to $c_0$ which is manifest from eq.(\ref{ihcondition}) and eq.(\ref{ihcondition1}). Thus, $c_0$ is no more than a gauge. Therefore, }no physical result should depend on the choice of $c_0$. { However,  unfortunately and as opposed to this expectation,} as we shall show shortly  that $c_0$ will appear in the symplectic structure of a spacetime admitting IH as its inner boundary { for real $\gamma$. Although the symplectic structure by itself is not a measurable quantity but we expect it to be gauge invariant as it explains the dynamics of the theory.} 
 
{ Now, let us consider one particular choice, namely, $~~c_0=1$. From eq.(\ref{ihcondition1}) it follows that}  $c_0=1\Rightarrow\spb{~\l}~=\spb{~\chi}$. {Further, considering the relation $c_0= (M\text{e})^{1/2}/\tilde c$ given in eq.(\ref{ihcondition1}), from eq.(\ref{ihcondition})} the choice $c_0=1$ leads to an independent equation in the present analysis:
 \begin{eqnarray}
\exp-\spb{~\alpha}=\frac{f(u)}{(M\text e)^{1/2}}.\label{redundantchoice}
 \end{eqnarray}
  Like $c_0=1$, choosing any other particular non-zero value of $c_0$ is equivalent to choosing a {\it fixed} relationship between the local Lorentz boost and the foliation of the IH {similar to eq.(\ref{redundantchoice})}. { Therefore, such a choice is redundant and no conclusions can be drawn from such choice regarding the laws of mechanics and surface gravity of the IH. In a nutshell, {\it such a choice can not be a `physical requirement'.}}

{ Nevertheless,}  we have identified the choice $c_0=1$ separately due to its importance {in the context of ref.\cite{per2} where one finds discussion regarding the symplectic structure of IH with real $SU(2)$ SAIB  variables. In ref.\cite{per2}, the choice $c_0=1$ was made by claiming it to be a `physical requirement' to obtain the correct value of surface gravity. Clearly, this is in sharp contrast to what we have explained. The issue will become even clearer shortly as we shall proceed through the investigations of the relevant equations that lead to the  symplectic structure.}

\section{Field equations in the canonical framework}\label{section2}
In this section we shall investigate the relevant equations in the canonical framework. The phase space variables are the SAIB connection and its conjugate momentum and the quantum theory is available only for real values of the Barbero-Immirzi parameter $(\gamma)$ \cite{ashlew,thiemann}.

\subsection{The Sen-Ashtekar-Immirzi-Barbero connection}

In the canonical framework an internal time-like vector $n^I:=e_\mu^In^\mu$
is chosen and kept fixed (see e.g. \cite{ashlew}),  $n^\mu$ is the unit time-like $(n^\mu n_\mu=-1)$ normal to the spatial slices.  
This choice only leaves the internal local rotational freedom which keep the chosen $n^I$ invariant \cite{ashlew}. Considering a fixed $n^I$, 
the Sen-Ashtekar-Immirzi-Barbero (SAIB) connection  is defined as
\begin{eqnarray}
A^i_\mu=\gamma\omega_\mu^{{\bf 0}i}-\frac{1}{2}\epsilon^{ijk}\omega_\mu^{jk}
\label{fv}
\end{eqnarray}
where $i={\bf 1,2,3}$ and $\gamma$ is called the Barbero-Immirzi parameter. $\omega^{IJ}_\mu$, with $I,J={\bf 0,1,2,3}$, are the components of the spin connection which can be written in terms of the tetrad components and the Christoffel symbols as follows
\begin{eqnarray}
\omega_\mu^{~IK}=-g^{\nu\alpha} e_\alpha^K\left(\partial_\mu e_\nu^I -\Gamma^\sigma_{\mu\nu}e_\sigma^I\right).\label{omegadef}
\end{eqnarray}
 The variable, defined in eq.(\ref{fv}), behaves as a connection while its pullback to a {\it spatial} slice is studied. However, it can be interpreted as a {\it spacetime} connection {\it only} for $\gamma=\pm i$ \cite{samuel} and in that case it is the complex Sen-Ashtekar connection.

\subsection{The equation on a 2-sphere}
Now, what are of importance in the present context  are the curvature of the SAIB connection defined as 
\begin{eqnarray}
F^i_{\mu\nu}=2\partial_{[\mu}A^i_{\nu]}+\epsilon^{ijk}A^j_{[\mu}A^k_{\nu]}\label{curvature}
\end{eqnarray}
and the following variable defined as 
\begin{eqnarray}
\Sigma^i_{\mu\nu}=2\epsilon^i_{jk}e_{[\mu}^je_{\nu]}^k. 
\end{eqnarray}
The study of the symplectic structure of a spacetime, admitting an IH as its inner boundary,  boils down to the task of finding the relation between these two variables correctly (see \cite{ih1,per2} for details). It should be noted that the pullbacks of $F$ and $\Sigma$ to any arbitrary 2-sphere  embedded in a four dimensional spacetime, to be denoted by $\underline F$ and $\underline \Sigma$ respectively, are related as \cite{per2}  
\begin{eqnarray}
\underline  F^i_{\mu\nu}=\left(\Psi_2-\Phi_{11}-\frac{R}{24}\right)\underline \Sigma^i_{\mu\nu}\label{gfi}
\end{eqnarray}
for $\gamma=\pm i$. Eq.(\ref{gfi}) is obtained by taking the pullback of a spacetime identity to an arbitrary 2-sphere (see \cite{ih1}) and then applying the partial gauge fixing (fixed $n^I$) on the internal space (see \cite{per2}). To mention, $R$ is the Ricci scalar associated with the spacetime and $
 \Psi_2 = C_{\mu\nu\alpha\beta}\ell^\mu m^\nu \bar m^\alpha n^\beta,
 \Phi_{11}= \frac{1}{4}R_{\mu\nu}(\ell^\mu n^\nu+m^\mu\bar{m}^\nu)
 $, 
are two Newman-Penrose scalars \cite{ih1}, 
where $C_{\mu\nu\alpha\beta}$ is the Weyl tensor and $R_{\mu\nu}$ is the Ricci tensor associated with the spacetime. In the present case, we have $\Phi_{11}=0=R$ and $\Psi_2=-2M/r^3$ (e.g. see \cite{chandra}). Therefore, eq.(\ref{gfi}) reduces to the following form
\begin{eqnarray}
\underline  F^i_{\mu\nu}=-\frac{2M}{r_0^3}~ \underline \Sigma^i_{\mu\nu}\label{gfis}
\end{eqnarray}
for $\gamma=\pm i$ and $r_0$ is the radius of the 2-sphere. Notably, this equation is not slicing dependent. 

Now, we aim to investigate the relation between $\underline F$ and $\underline\Sigma$ both for real and imaginary values of $\gamma$. To do that we calculate the components of $\underline F$ and $\underline{\Sigma}$ separately. From eq.(\ref{omegadef}), the non-zero components of the spin connection can be calculated to be 
\begin{eqnarray}
&&\omega_t^{~{\bf 01}}=\frac{\Omega'}{2\Omega}-\dot\alpha, ~~~~~\omega_x^{~{\bf 01}}=\frac{\dot\Omega}{2\Omega}-\alpha', ~~~~~\omega_\theta^{~{\bf 02}}=\Omega^{-1/2}(\dot r\cosh\alpha-r'\sinh\alpha),\nn\\
&&\omega_\theta^{~{\bf 12}}=\Omega^{-1/2}(\dot r\sinh\alpha-r'\cosh\alpha),~~~~~ \omega_\phi^{~{\bf 03}}=\frac{\sin\theta}{\Omega^{1/2}}(\dot r\cosh\alpha-r'\sinh\alpha),\nn\\
&&\omega_\phi^{~{\bf 13}}=\frac{\sin\theta}{\Omega^{1/2}}(\dot r\sinh\alpha-r'\cosh\alpha),~~~~\omega_\phi^{~{\bf 23}}=-\cos\theta
\end{eqnarray}
and the corresponding anti-symmetric ones with respect to the internal indices. Then using the definition given in eq.(\ref{fv}), one can find the  non-zero components of the SAIB connection to be 
\begin{eqnarray}
&&A_t^{\bf 1}=\gamma\left(\frac{\Omega'}{2\Omega}-\dot\alpha\right),
~~~~~A_x^{\bf 1}=\gamma\left(\frac{\dot\Omega}{2\Omega}-\alpha'\right),~~~~~A^{\bf 1}_\phi=\cos\theta,\nn\\
&&A_\theta^{\bf 2}=\frac{\gamma}{\Omega^{1/2}}\left(\dot r\cosh\alpha-r'\sinh\alpha\right),~~~~~A^{\bf 2}_\phi=\frac{\sin\theta}{\Omega^{1/2}}(\dot r \sinh\alpha-r'\cosh\alpha),\nn\\
&&A_\theta^{\bf 3}=\frac{1}{\Omega^{1/2}}\left( r' \cosh\alpha-\dot r\sinh\alpha\right),~~~~~A^{\bf 3}_\phi=\frac{\gamma\sin\theta}{\Omega^{1/2}}(\dot r \cosh\alpha-r'\sinh\alpha).\label{conncompo}
\end{eqnarray}
Then one can calculate $\underline F$ by using the definition in eq.(\ref{curvature}). The non-zero components are given by:
\begin{eqnarray}
\underline F^{\bf 1}_{~\th\phi}&=&-\sin\theta\left[\frac{2M}{r_0}-(1+\gamma^2)\frac{\exp[-r_0/2M]}{r_0}\left(t_0\cosh\underline{\alpha}+x_0\sinh\underline{\alpha}\right)^2\right]\\
\underline F^{\bf 2}_{~\th\phi}&=&~~\frac{\exp[-r_0/4M]}{r_0^{1/2}}\left[-\sin\theta \left(t_0\cosh\underline{\alpha}+x_0\sinh\underline{\alpha}\right)\left(\partial_{\theta}\underline{\alpha}\right) +\gamma\left(t_0\sinh\underline{\alpha}+x_0\cosh\underline{\alpha}\right)\left(\partial_{\phi}\underline{\alpha}\right)\right]\\
\underline F^{\bf 3}_{~\th\phi}&=&-\frac{\exp[-r_0/4M]}{r_0^{1/2}}\left[\gamma\sin\theta \left(t_0\sinh\underline{\alpha}+x_0\cosh\underline{\alpha}\right)\left(\partial_{\theta}\underline{\alpha}\right) +\left(t_0\cosh\underline{\alpha}+x_0\sinh\underline{\alpha}\right)\left(\partial_{\phi}\underline{\alpha}\right)\right]
\end{eqnarray}
 $t_0$ and $x_0$ are coordinates of the particular 2-sphere under consideration in the $t-x$ plane and hence, they are related as
\begin{eqnarray}
&&t_0^2-x_0^2=-(r_0-2M)e^{r_0/2M}.
\end{eqnarray}
On the other hand, one can calculate that the only non-vanishing component of $\underline{\Sigma}$ is
\begin{eqnarray}
\underline \Sigma^{\bf 1}_{\theta\phi}=2r_0^2\sin\theta
\end{eqnarray} 
We may note that since $\underline \Sigma^{\bf 2}_{\theta\phi}$ and $\underline \Sigma^{\bf 3}_{\theta\phi}$ vanish identically, it can be concluded that the boost parameter $\alpha$ in the $t-x$ plane needs to be spherically symmetric (i.e. $\partial_\theta \alpha=0=\partial_\phi \alpha$) so that eq.(\ref{gfis}) holds for $\gamma=\pm i$. Having said this, for $\gamma\neq\pm i$, we have the following equation at hand:
\begin{eqnarray}
\underline F^{i}_{~\mu\nu}&=&-\frac{1}{r_0^2}\left[\frac{2M}{r_0}-(1+\gamma^2)\frac{\exp[-r_0/2M]}{r_0}\left(t_0\cosh{\underline \alpha}+x_0\sinh{\underline \alpha}\right)^2\right]\underline \Sigma^{i}_{~\mu\nu}\label{2sphere}
\end{eqnarray}
which has the property that the proportionality factor between $\underline F$ and $\underline \Sigma$ depends on the slicing of the spacetime 
and on the internal gauge choice. 
This is a fundamental inconsistency, reflecting the fact that the SAIB connection variable does not have a spacetime interpretation for $\gamma\neq\pm i$ \cite{samuel}. Hence, one may wonder why we should be interested in the relation between $\underline{F}$ and $\underline{\Sigma}$ at the first place. The reason is that this relation plays the fundamental role in writing down the symplectic structure of a spacetime admitting an IH as its inner boundary \cite{ih1,per2}.

If one wants to avoid the above inconsistency, one is forced to introduce a parameter of unusual nature, say $\sigma$,  which characterizes a relation between the internal boost parameter and the slicing of the spacetime such that
\begin{eqnarray}
\underline F^{i}_{~\mu\nu}&=&-\frac{2M}{r_0^3}\left[1-(1+\gamma^2)\sigma^2\right]\underline \Sigma^{i}_{~\mu\nu}\label{2spherefix}
\end{eqnarray}
with 
\begin{eqnarray}
\sigma^2:=\frac{\exp[-r_0/2M]}{2M}\left(t_0\cosh{\underline \alpha}+x_0\sinh{\underline \alpha}\right)^2.
\end{eqnarray}
$\sigma$ can be any non-zero real number satisfying the condition $\sigma^2\neq(1+\gamma^2)^{-1}$.
 Doing this, one actually restricts the local Lorentz group to a subgroup whose boosts are tailored in a particular fashion that is dependent on the slicing of the spacetime. So, the `local' boosts have to know about the `global' slicing. Therefore, $\sigma$ marks different sectors of the boost parameter within which the equations on the 2-sphere are slicing independent.
 
In general, one can choose $\sigma$ to be dependent on spacetime coordinates and introduce a new gauge field. However, we keep that possibility out of the present discussion and restrict $\sigma$ to be just an arbitrary constant parameter (independent of $\gamma$).

\subsection{On the isolated horizon}
Let us see what the scenario is, if we concentrate on the IH only. We shall replace now the `underline' with `left double arrow' to indicate that the 2-sphere is a cross-section of the IH i.e. $r_0=2M$ and $x_0=t_0=f(u_0)$. Then,  eq.(\ref{2sphere}) reduces to the following form:
\begin{eqnarray}
\sdpb{F^{i}}_{\mu\nu}&=&-\frac{2\pi}{A_{IH}}\left[1-(1+\gamma^2)\sigma^2\right]\sdpb{\Sigma^{i}}_{\mu\nu}\label{slice}
\end{eqnarray}
where we have used the fact that $A_{IH}=16\pi M^2$ is the area of cross-section of the IH and we have 
\begin{eqnarray}
\sigma=\frac{f(u_0)}{(2M\text{e})^{1/2}}~\exp ~\sdpb{~\alpha}\label{sigma}
\end{eqnarray} 
on the IH. One can see that eq.(\ref{sigma}) is just the pullback of eq.(\ref{ihcondition}) on a slice of the IH at $u_0$, with $\tilde c=\sigma(2M\text{e})^{1/2}$ and therefore, from eq.(\ref{ihcondition1}) we have $c_0=1/\sigma\sqrt{2}$ implying $\spb{~\ell}=\frac{1}{\sigma\sqrt 2}~\spb{~\chi}$. It is now clear that the parameter $\sigma$ that appears in eq.(\ref{slice}) is actually the representative of the equivalence class of null generators of the IH and we have seen that  the laws of mechanics of the IH and the value of surface gravity does not depend on the choice of $\sigma$.   Using eq.(\ref{slice}), one can show that the symplectic structure of the spacetime with IH as an inner boundary is given by
\begin{eqnarray}
32\pi \gamma\bold{\Omega}(\delta_1,\delta_2)=\int_M 2~\delta_{[1}\Sigma^i\wedge \delta_{2]}A_i\underbrace{~-~\frac{A_{IH}}{2\pi[1-(1+\gamma^2)\sigma^2]}\int_{S_M}\delta_{[1}{\cal A}^i\wedge \delta_{2]}{\cal A}_i}_{\text{ SS}}~,\label{symplectic}
\end{eqnarray}
where $M$ is a spatial slice `sometime' between $\Phi_i$ and $\Phi_f$, $S_M$ is the intersection 2-sphere of $M$ with the IH and ${\cal A}^i:=\sdpb{~A}^i$. One can see ref.\cite{per2} for the details of the derivation, but with the specific {\it choice} $\sigma=1/\sqrt 2$. { Considering the fact that $c_0=1/\sigma\sqrt2$, the choice $\sigma=1/\sqrt 2$ implies $c_0=1$. As we have already discussed in subsection \ref{subred}, the choice $c_0=1$ or equivalently $\sigma=1/\sqrt 2$ is redundant  and certainly not any physical requirement. Therefore, eq.(\ref{symplectic}), retains a dependence on the parameter $\sigma$ that represents the rescaling freedom of the null generators of the IH. We hope that now it is  completely clear why we discussed separately the choice $c_0=1$ and emphatically called it redundant.}

Two remarks need to be made about the result in eq.(\ref{symplectic}).
\begin{itemize}
\item $\sigma$ does not appear at all in the symplectic structure for $\gamma=\pm i$, which is an implication of the fact that the SAIB variable has an interpretation of a spacetime gauge field only for $\gamma=\pm i$. 
\item Considering only real values of $\gamma$, the symplectic structure diverges for $\sigma^2=(1+\gamma^2)^{-1}$. This implies that one member of the equivalence class of null generators of the IH  becomes `special', which implies a violation of the rescaling symmetry of the null generators of the IH.
\end{itemize} 
  Nonetheless,  a viable quantum theory is available only for real and positive values of $\gamma$. Therefore, it is founded on a flawed classical setup.  {Thus, in order to proceed with any sort of calculation, one has to bear with this unavoidable symmetry violation and consider omitting, by hand,  the boosted frame on the IH given by $\sigma^2=(1+\gamma^2)^{-1}$.  Otherwise, one has to discard the theory at the classical level itself, for real $SU(2)$ SAIB variable.}

 \subsubsection{The celebrated equation}
 {In this short subsection, let us offer some comments on a particular equation that has played crucial role in the development of the related contemporary literature.}
 We think, at this point, it is highly important to mention that eq.(\ref{slice}), with the particular choice $\sigma=1/\sqrt 2$, results in the following equation:
 \begin{eqnarray}
 \sdpb{F^{i}}_{\mu\nu}&=&-\frac{\pi}{A_{IH}}\left(1-\gamma^2\right)\sdpb{\Sigma^{i}}_{\mu\nu}\label{slice1}.
 \end{eqnarray}
 Eq.(\ref{slice1}) has appeared frequently in the literature till date \cite{perezcite} and has played pivotal role on several occasions for further build up of new results (which is the reason for calling the equation as ``celebrated'').  Needless to say, any  result or any conclusion that is based on eq.(\ref{slice1}), is gauge dependent in the sense that eq.(\ref{slice1}) results from a particular gauge choice. { The reason for isolating eq.(\ref{slice1}) is to emphasize the basic flaw regarding gauge fixing that has percolated through a non-negligible part of the literature which blindly rely on eq.(\ref{slice1}).}


\section{Remarks on entropy calculation} \label{remarks}

{ As it is now clear from the above discussion that the rescaling symmetry of an IH, real $SU(2)$ SAIB variable (i.e. real value of $\gamma$) and hence, the quantization program, do not tally with each other.  Nevertheless,} we  shed some light on the role  played by the parameter $\sigma$ in the existing literature concerning entropy calculation of $SU(2)$ IH, { with an assumption that $\gamma$ has been experimentally determined to be $\gamma_{e}$ (say). In the discussion we exclude the condition $\sigma^2=(1+\gamma^2)^{-1}$ and discuss the logical viability of the calculations which exist in literature from a field theoretic viewpoint.}

Let us point out that eq.(\ref{slice}) can be cast as the equation of motion for a Chern-Simons theory coupled to a source and this equation holds the key for the entropy calculation \cite{per2,per3,sigma,km11}. In fact, in the quantum theory, the microstates of the IH belong to the Hilbert space of a quantum Chern-Simons theory coupled to point-like sources carrying quantum numbers that can take values like $1/2,1,\cdots,k/2$ where $k$ is called the level of the Chern-Simons theory \cite{qg1,qg2,per2,per3,sigma}. The number of microstates corresponding to the Hilbert space of a Chern-Simons theory with level $k$, is some function $\Theta(k)$ \cite{ampm}.  Having said this,  { let us clarify why, among the two existing view points -- one adopted in refs.\cite{sigma,km11} and the other in ref.\cite{per4} -- the former is logically viable but  the later is not. In the process, we also point out why ref.\cite{per2} and ref.\cite{per4} present  self-contradictory research investigations\footnote{We call it ``self-contradictory'' because two authors are common in ref.\cite{per2} and ref.\cite{per4}.}.} 

\subsection{The Kaul-Majumdar (KM) scenario}
In \cite{sigma,km11} it has been rigorously argued that the $\sigma$-dependence should be used to define the source term and {\it not} the Chern-Simons level because it affects the entropy  which is a physical quantity. Therefore, one can view eq.(\ref{symplectic}) as an `equivalence class of symplectic structures' corresponding to the equivalence class of null generators that lead to the same physics associated with the IH viz. mechanical laws, value of surface gravity and the resultant entropy. Notably, the source term vanishes for $\sigma^2= (1+\gamma^2)^{-1}$ \cite{km11,sigma}, which we have removed `by hand' from consideration at the beginning. This approach has the following highlighted features:
\begin{itemize}
	\item The Chern-Simons level $k:=A_{IH}/4\pi\gamma\lp^2$ where $\lp$ is the Planck length. Therefore, $A_{IH}/\lp^2\gg 1 \implies k\gg 1$. 
	\item The Boltzmann entropy formula  is used as the starting point i.e. $S=\ln \Theta(k)=\ln \Theta(A_{IH}/\lp^2,\gamma)$.
	\item One needs to choose a particular value of $\gamma$, say $\gamma_0$,  to get the area law i.e. $S=A_{IH}/4\lp^2$. 
	\item $\sigma$ does not affect the entropy calculation and remains arbitrary implying that all the members of the equivalence class of null generators lead to the same physical result. 
	\item The theory is falsifiable if we find $\gamma_e\neq \gamma_0$.
\end{itemize}
{ Thus, the KM scenario seems to be logically viable from a field theoretic perspective where the coupling constant remains unambiguous and it is verifiable by experiment as one expects any theory to be.}

\subsection{The Perez-Pranzetti (PP) scenario}
In \cite{per4}, the Chern-Simons level is defined in a $\sigma$-dependent way. In that case, demanding the entropy to be given by the area law, one finds a relation between $\sigma$, $\gamma$ and $A_{IH}$. Therefore, for a given area black hole, the entropy is given by the area law if $\sigma$ and $\gamma$ has a particular relationship. This seems to be elegant because one obtains the area law for any $\gamma$ that is fixed up to a relation with $\sigma$, an arbitrary positive definite parameter. However, the problem arises if we use $\gamma=\gamma_e$ in the results of \cite{per4}.  Then, what one is left with is a relation between $\sigma$ and $A_{IH}$. This implies that, for a given $A_{IH}$, there is a preferred null generator, among the equivalence class of null generators of the IH, that leads to the physical results. It  also implies a fixed relationship between the local Lorentz boost and the foliation of the IH. This explicitly breaks the intrinsic symmetries of the IH even after considering $\sigma^2\neq (1+\gamma^2)^{-1}$. We highlight the main features of this approach of entropy calculation:
\begin{itemize}
	\item The Chern-Simons level $k:=A_{IH}/8\pi\gamma(1-\bar\gamma^2)\lp^2$, where $\bar\gamma^2:=(1+\gamma^2)\sigma^2$. Therefore, $A_{IH}/\lp^2\gg 1$ does not imply $k\gg 1$ and $k$ is now an  independent input just because of the arbitrary parameter $\sigma$.
	\item The Boltzmann entropy formula is used as the starting point, as usual, leading to $S=\ln \Theta(k)=\ln \Theta(A_{IH}/\lp^2,\gamma,\sigma)$.
	\item To get the area law, the following relation has to be satisfied i.e. $\ln \Theta(A_{IH}/\lp^2,\gamma,\sigma)=A_{IH}/4\lp^2$. Therefore, for any given $A_{IH}$, we only have a relation between $\sigma$ and $\gamma$. We need not choose $\gamma$.  
	\item If we put $\gamma=\gamma_e$, then, for a given $A_{IH}$ we have a fixed $\sigma$, implying the violation of the symmetry of the IH even after considering $\sigma^2\neq (1+\gamma^2)^{-1}$. 
\end{itemize}
It may be noted that in \cite{per4} no particular choice of the parameter $\sigma$ has been made, although it was already shown in \cite{per2} (albeit incorrectly, as we have pointed out) that $\sigma$ needs to be chosen. Therefore, there is a serious inconsistency that exists in part of the literature concerning the entropy calculation of $SU(2)$ IH created due to self-contradictory research investigations made in ref. \cite{per2} and ref. \cite{per4}. { One can not take ref.\cite{per4} as an amendment to ref.\cite{per2} because the problems of using surface gravity as an input was not even addressed in ref.\cite{per4}.} { Moreover, the scenario presented in ref.\cite{per4} is, at the least, logically impracticable as it worsens the problem by being suggestive of the fact that there should be a preferred Lorentz boosted frame on the horizon for a given  area, as $\sigma$ becomes related to $A_{IH}$, if the value of $\gamma$ is experimentally determined, even after considering $\sigma^2\neq (1+\gamma^2)^{-1}$.}

\section{Conclusion}\label{section3}
{ Let us summarize the facts investigated in this work.} We have verified that the symplectic structure of a spacetime admitting an $SU(2)$ IH as an inner boundary, { particularly the contribution from the IH that is referred to as SS in eq.(\ref{symplectic})}, written in terms of the  Sen-Ashtekar-Immirzi-Barbero (SAIB) variables, is dependent on a positive definite parameter $(\sigma)$ that represents the equivalence class of null generators of the IH. The mechanical laws and the value of surface gravity associated with an IH do not depend on any particular value of $\sigma$. Therefore, there is no physical motivation behind a choice of $\sigma$, as opposed to the claim made in ref. \cite{per2} that lay the foundations of the theory of $SU(2)$ IH. { The importance of these finding becomes  profoundly manifest while one finds the impact of ref.\cite{per2} in the associated literature till date \cite{perezcite}.} 
 
  The  $\sigma$-dependence of the SS disappears when the Barbero-Immirzi parameter, $\gamma$, is equal to $\pm i$. The SS diverges for $\sigma^2= (1+\gamma^2)^{-1}$, thus implying that the full symmetry of the IH does not lead to a well defined  SS. These results are  direct manifestations of the fact that the Sen-Ashtekar-Immirzi-Barbero connection, which is defined on a spatial slice, can be interpreted as the pullback of a spacetime connection only for $\gamma=\pm i$ \cite{samuel}. However, a viable quantum theory of $SU(2)$ IH exists only for real values of $\gamma$. Therefore, such a theory is founded on a flawed classical setup.
  
  { To put it simply, even before going into the quantization, the classical setup is devoid of explanations of the laws of mechanics and the construction of the IH framework, along with its symmetries, for the real $SU(2)$ SAIB variable (real $\gamma$). Thus, the question arises that if we can not make sense of black hole thermodynamics with these variables then what does it even mean to talk about `surface gravity', `entropy', etc. Such a setup is, at the least, devoid of the motivation for quantization and entropy calculation in the first place.}
  
  { We note further that, if such a flaw is ignored nonetheless, then one can proceed with the calculations to find the  microstates and entropy from a purely field theoretic perspective, but only by imposing a restriction  $\sigma^2\neq(1+\gamma^2)^{-1}$ `by hand'. In such case, two different viewpoints exist in the literature, namely, in refs.\cite{km11,sigma} (KM scenario) and in ref.\cite{per4} (PP scenario). We point out that, if $\gamma$ were experimentally determined to have a fixed numerical value, then the two scenarios lead to two different conclusions. In the KM scenario, the theory is falsifiable if the chosen   value of $\gamma$ does not match the experimental value. In the PP scenario, an experimentally fitted value of $\gamma$ implicitly suggests that for a given area there is a preferred Lorentz boosted frame of the horizon even for $\sigma^2\neq(1+\gamma^2)^{-1}$. Therefore, the KM scenario is logically viable but the PP scenario does not seem to be so.}
  
  { Thus, considering the basic nature of the issues that we have investigated in this work, on logical grounds it seems to us that, as far as real $SU(2)$ SAIB variable is concerned, the basic construction of the IH horizon framework leading to the laws of mechanics (hence, the definition of surface gravity) needs to be sorted out before one even thinks about quantization and entropy of the IH with such a variable. Only the performance  of such a task can provide a clear justification of using such a variable, consequently establishing an unobjectionable motivation for quantization of the classical theory in the LQG framework.}

\vspace{0.3cm}

{\bf Acknowledgement:} The author thanks  Romesh Kaul and Parthasarathi Majumdar for helpful discussions. Later stage of this work is supported by the Department of Science and Technology of India through the INSPIRE Faculty Fellowship, Grant no.- IFA18-PH208.

\end{document}